\documentclass[aps,twocolumn,prl,showpacs,floatfix,superscriptaddress]{revtex4}
\usepackage{graphicx}
\usepackage{amsmath}
\usepackage{amsfonts}
\usepackage{amssymb}
\newcommand{\na}{{Na$_2$IrO$_3$}}

\begin{document}
\title{Na$_{2}$IrO$_{3}$ as a molecular orbital crystal}
\author{I. I. Mazin}
\affiliation{Code 6393, Naval Research Laboratory, Washington, DC 20375, USA}

\author{Harald O. Jeschke}
\affiliation{Institut f\"ur Theoretische Physik, Goethe-Universit\"at Frankfurt, 60438 Frankfurt am Main, Germany}

\author{Kateryna Foyevtsova}
\affiliation{Institut f\"ur Theoretische Physik, Goethe-Universit\"at Frankfurt, 60438 Frankfurt am Main, Germany}

\author{Roser Valent\'\i}
\affiliation{Institut f\"ur Theoretische Physik, Goethe-Universit\"at Frankfurt, 60438 Frankfurt am Main, Germany}

\author{D. I. Khomskii}
\affiliation{II. Physikalisches Institut, Universit\"at zu K\"oln, Z\"ulpicher Stra{\ss}e 77, 50937 K\"oln, Germany}
\date{\today }

\pacs{75.10.-b,75.10.Jm,71.70.Ej,71.15.Mb}

\begin{abstract}
Contrary to previous studies that classify {\na} as a realization of
the Heisenberg-Kitaev model with dominant spin-orbit coupling, we show
that this system represents a highly unusual case in which the
electronic structure is dominated by the formation of quasi-molecular
orbitals (QMOs), with substantial quenching of the orbital
moments. The QMOs consist of six atomic orbitals on an Ir hexagon, but
each Ir atom belongs to three different QMOs. The concept of such QMOs
in solids invokes very different physics compared to the models
considered previously.  Employing density functional theory
calculations and model considerations we find that both the insulating
behavior and the experimentally observed zigzag antiferromagnetism in
{\na} naturally follow from the QMO model.
\end{abstract}
\maketitle

High interest in the recently synthesized hexagonal
iridates~\cite{felner,kobayashi,gegenwart} is due to the hypothesis
~\cite{Jackeli2009,Shitade2009} that the electronic structure in these
materials is dominated by the spin-orbit (SO) interaction. In this
case, the Ir $t_{2g}$ bands are most naturally described by
relativistic atomic orbitals with the effective angular moment,
$j_{\rm eff}=3/2$ and $j_{\rm eff}=1/2.$ In this approximation, the
splitting between the 3/2 and 1/2 states is larger than their
dispersion. The upper band $j_{\rm eff}=1/2$ is half filled and Ir
atoms can be described as localized ($j_{\rm eff}=1/2,$ $M=1$
$\mu_{\rm B}$) magnetic moments~\cite{note1} with the exchange
interaction strongly affected by SO coupling.  In particular, this
picture leads to a very appealing framework known as Heisenberg-Kitaev
model~\cite{jackeli,trebst}, with highly nontrivial physical
properties. However, experimental evidence for the $j_{\rm eff}$
scenario is lacking~\cite{jeff}.

In this Letter, based on \textit{ab initio} density functional theory
(DFT) calculations and model considerations, we show that this picture
does not apply to the actual {\na}.  Instead, this system represents a
highly unusual case where the formation of electronic structure is
dominated by quasi-molecular orbitals (QMOs), which involve six Ir
atoms arranged in a hexagon. What distinguishes this picture from
molecular solids is that there is no associated spatial
clusterization, but each Ir atom (via its three $t_{2g}$ orbitals)
participates in three different QMOs, yet in the first approximation
there is no inter-QMO hopping the thus formed bands are
dispersionless.

Such an electronic structure calls for a new approach. There is no
known recipe for handling its magnetic properties, or adding Coulomb
correlations, for instance. While we will not present a complete
theory of spin dynamics and correlations in the QMO framework, we will
outline the general directions and most important questions, in the
expectation that this will stimulate more theoretical and experimental
work and eventually generate more insight. Yet, the key observable
features of {\na}: small magnetic moment, unusual zigzag
antiferromagnetism, and Mott-enhanced insulating behavior, are
naturally consistent with the QMO framework.

The main crystallographic element of {\na} (see SI) is an Ir$^{4+}$
(5$d^{5}$) honeycomb layer with a Na$^{1+}$ ion located at its
center. Each Ir is surrounded by an O octahedron, squeezed along the
cubic [111] (hexagonal $z$) axis. Therefore, Ir $d$-states are split
into an upper $e_{g}$ doublet and a lower $t_{2g}$ triplet. The [111]
squeezing further splits the $t_{2g}$ levels into a doublet and
singlet; initially this effect was
neglected~\cite{Jackeli2009,jackeli,trebst}, however, it was later
included~\cite{KimKim,trig1} (and overestimated) to explain the
observed deviations from the Heisenberg-Kitaev model.

In the previous works, after identifying the $t_{2g}-e_{g}$ splitting
it was assumed that the energy scales are $W$ $<$ ($J_{H}$, $\lambda$)
$<$ $U$, where $W\sim4t$ is the $d$-electron band width, $t$ the
effective hopping parameter, $J_{H}$ the Hund's rule coupling,
$\lambda$ the SO parameter, and $U$ the on-site Coulomb repulsion. In
this limit, the electrons are localized and the system is a Mott
insulator.  While $\lambda$ $\sim$ 0.4-0.5 eV for $5d$ ions, the
bandwidth for 5$d$ orbitals is 1.5-2 eV and $U\sim1-2$ eV,
$J_{H}\sim0.5$ eV, reduced compared to typical $U\sim 3-5$ eV and
$J_{H}\sim0.8-0.9$ eV for $3d$ electrons. Many-body renormalization
may narrow the bands by a factor $(m^{\ast}/m);$ however, given that
in Ir $U\sim W$, it is unrealistic to expect a large renormalization.
Therefore, the usual starting point $W< (J_{H},\lambda)<U$ is not
valid here, rather, the system is close to an itinerant
regime. $I.e.$, one cannot justify reducing the description of {\na}
(and possibly other iridates) to an effective $j=1/2$ model, decoupled
from the other $j_{eff}$ states.

\begin{figure}[tbh]
\begin{center}
\includegraphics[angle=-90,width=.95 \columnwidth]{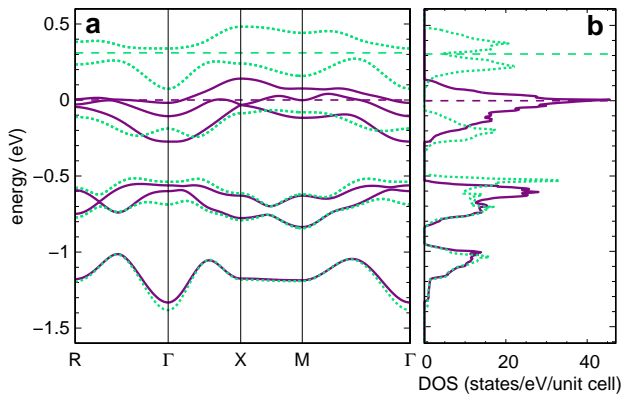}
\end{center}
\caption{(color online) Electronic structure of the non-magnetic {\na}
  for the experimentally determined~\cite{Radu} crystal structure. The
  calculations were performed with the full potential local orbital
  (FPLO) basis using the generalized gradient approximation (see SI).
  The solid purple and dotted green lines refer to calculations
  without and with SO interaction, respectively. Note that the Fermi
  levels (shown by the horizontal dotted lines) are not aligned.  }
\label{bandSO}
\end{figure}

\begin{figure}[tbh]
\caption{(color online) Most relevant O $p$-assisted hopping paths in
  idealized {\na} structure.  For each of the three Ir-Ir bond types
  only hopping between two particular $t_{2g}$ orbitals is
  possible. The same holds for the second and third nearest neighbor
  hopping \textit{via} O $p$ and Na $s$ orbitals. Ir-Ir bonds are
  color coded as follows: $xy$ bonds are shown by blue lines, $xz$
  bonds by green, and $yz$ bonds by red ones.}
\label{fig:structure2}
\begin{center}
\includegraphics[angle=-90,width=0.9 \columnwidth]{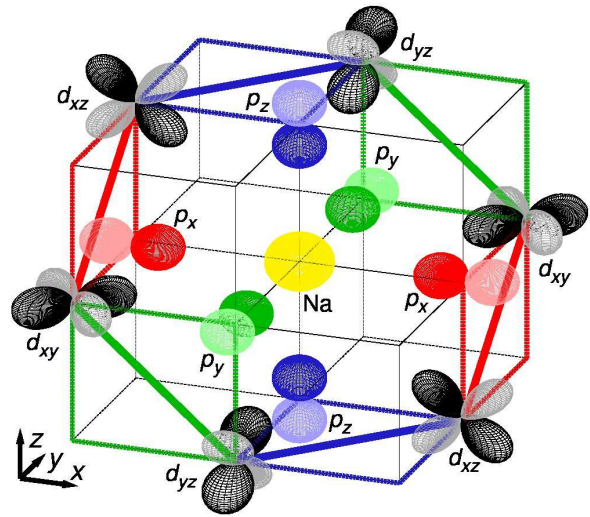}
\end{center}
\end{figure}

Thus, the first step (usually skipped) is to understand the
non-relativistic band structure. We have therefore performed DFT
calculations (see SI) initially without SO effects (see
Fig.~\ref{bandSO}, solid purple lines).  Inverting the band structure
results (see SI), we obtained the corresponding tight-binding
Hamiltonian.  The leading channel (by far) is the nearest neighbor
(NN) O-assisted hopping between unlike orbitals (see
Fig.~\ref{fig:structure2}). This was also correctly identified
previously~\cite{Jackeli2009,Shitade2009}. There are three different
types of NN Ir-Ir bonds; for one (we name it $xy$ bond) (see Fig. 3)
this hopping is only allowed between $d_{xz}$ and $d_{yz}$ orbitals,
for the next ($xz$) between $d_{yz}$ and $d_{xy}$ orbitals and for the
third bond ($yz$) between $d_{xy}$ and $d_{xz}$. In our calculations
this hopping, $t_{1}^{\prime}$ (the prime indicates that the hopping
is $via$ O) is about 270 meV. Perturbatively, this term is
proportional to $t_{pd\pi}^{2}/(E_{t_{2g}}-E_{p}),$ where $p$ stands
for the O $p$ states. Ref.~\cite{Shitade2009} pointed out another
(next nearest neighbors, NNN) O-assisted term, which we find to be
$\sim 75$ meV. Jackeli and Khalliulin~\cite{Jackeli2009} invoked
another NN hopping process, between like orbitals pointing directly to
each other.  Despite the short Ir-Ir distance, these matrix elements
are surprisingly small, $\lesssim30$ meV. Finally, some
authors~\cite{KimKim,trig1} addressed the trigonal squeeze, which
creates non-zero matrix elements between the same-site $t_{2g}$
orbitals.

The main feature of the calculated non-relativistic band structure
(see Fig.~\ref{bandSO}) is formation of a singly degenerate (not
counting spins) band state at $\sim -1.2$ eV, a doubly degenerate one
at $-0.7$ eV, and a three-band manifold between $-0.3$ and $0.2$
eV. This clear separation, of the order of 0.3 eV, cannot be related
to the trigonal squeeze, as this can only split the 6 $t_{2g}$ bands
(there are two Ir per cell) into a doublet and quartet.

\begin{figure}[tbh]
\begin{center}
\includegraphics[width=0.9 \columnwidth]{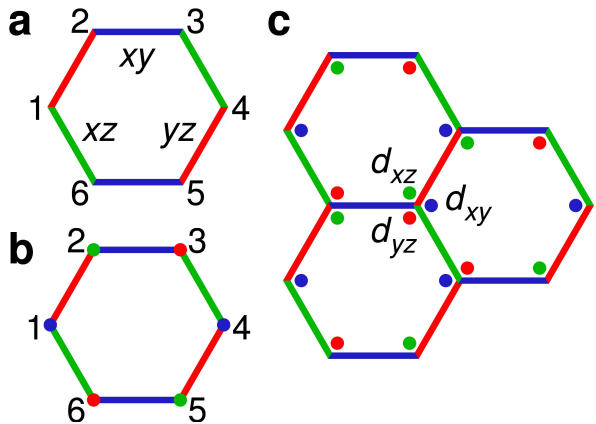}
\end{center}
\caption{(color online) (a) Schematic plot of a Ir$_{6}$Na hexagon. We
  use the same color coding as in Fig.~\ref{fig:structure2}, $xy$
  bonds are shown by blue lines and $d_{xy}$ orbitals by blue dots,
  \textit{etc.} (b) A quasi-molecular composite orbital on a given
  hexagon. (c) Three neighboring quasi-molecular orbitals.}
\label{qmo}
\end{figure}

In order to understand this, we start with the dominant hopping, the
NN O-assisted $t_{1}^{\prime}$. Let us consider an electron on a given
Ir site in a particular orbital state, say, $d_{xz}.$ The site has
three NN neighbors. As discussed above, this electron can hop, with
the amplitude $t_{1}^{\prime},$ to a neighboring state of $d_{yz}$
symmetry, located at a particular NN site. From there, it can hop
further into a $d_{xy}$ state on the next site, and so on (see
Figs.~\ref{fig:structure2} and ~\ref{qmo}).  At each site, the
electron has only one bond along which it can hop. Following the
electron around, we see that after six hops it returns to the same
state and site from where it started. \textit{This means that in the
  NN }$t_{1}^{\prime}$ \textit{approximation every electron is fully
  localized within 6 sites forming a hexagon.} Such a state could be
called a molecular orbital, except that there are no spatially
separated molecules on which electrons are localized.  Each Ir belongs
to three hexagons, and each Ir-Ir bond to two. Thus, three different
$t_{2g}$ orbitals on each Ir site belong to three different
\textquotedblleft quasi-molecular\textquotedblright\ orbitals (QMO)
and these QMOs are fully localized in this approximation
(Fig.~\ref{qmo}).

Six QMOs localized on a particular hexagon form six levels, listed in
Table~\ref{table}, grouped into the lowest B$_{1u}$ singlet, the
highest A$_{1g}$ singlet, and two doublets E$_{1g}$ and E$_{2u}$. The
energy separation between the lowest and the highest level is
$4t_{1}^{\prime},$ which is close to the calculated total
non-relativistic $t_{2g}$ band width.
\begin{table}[ptb]
\caption{Six quasi-molecular orbitals formed by the six $t_{2g}$
  atomic orbitals on a hexagon. $(\omega=\exp(i\pi/3))$.  Note that
  $t_{1}^{\prime}$$>$ 0 and $t_{2}^{\prime}$$<$0}
\label{table}
\begin{center}
\begin{tabular}
[c]{l|l|l}\hline
Symmetry & Eigenenergy & Eigenvector(s)\\\hline
$A_{1g}$ & $2(t_{1}^{\prime} + t_{2}^{\prime})$ & $(1,1,1,1,1,1)$ \\\hline
$E_{2u}$ & $t_{1}^{\prime} - t_{2}^{\prime}$ & $(1, \omega, \omega^{2},
-1, \omega^{4} , \omega^{5})$\\
(twofold) &  & $(1, \omega^{5}, \omega^{4}, -1, \omega^{2}, \omega
)$\\\hline
$E_{1g}$ & $-t_{1}^{\prime} - t_{2}^{\prime}$ & $(1, \omega^{2}, \omega
^{4}, 1, \omega^{2}, \omega^{4})$\\
(twofold) &  & $(1, \omega^{4}, \omega^{2}, 1, \omega^{4},
\omega^{2})$\\\hline
$B_{1u}$ & $-2(t_{1}^{\prime} + t_{2}^{\prime})$ & $(1,-1,1,-1,1,-1)$
\end{tabular}
\end{center}
\end{table}

We now add the O-assisted NNN hopping $t_{2}^{\prime}$. Here there are
several such paths. However, the dominant hopping takes advantage of
the diffuse Na $s$ orbital (see Fig.~\ref{fig:structure2}), and is
proportional to
$t_{pd\pi}^{2}t_{sp}^{2}/(E_{t_{2g}}-E_{p})^{2}(E_{t_{2g}}-E_{s})<0$. It
connects unlike NNN $t_{2g}$ orbitals that belong to the same QMO, and
therefore retains the complete localization of individual QMOs. It
does shift the energy levels though, as shown in
Table~\ref{table}. The upper singlet and doublet get closer and the
lower bands move apart providing the average energy separations of
$\sim 0.5$, $\sim 0.6$, and $\sim 0.1$~eV among the calculated
non-relativistic subbands (at $|t_{1}^{\prime}/t_{2}^{\prime}|=2$ the
upper two levels merge; in reality,
$|t_{1}^{\prime}/t_{2}^{\prime}|\approx 3.3)$. Given that the subband
widths are 0.2--0.3 eV, obviously, the upper doublet and singlet merge
to form one three-band manifold.

Several effects contribute to the residual dispersion of the QMO
subbands. The trigonal splitting plays a role, albeit smaller than
often assumed: the trigonal hybridization is $\Delta\approx25$ meV
(the splitting being $3\Delta$).  This may seem surprising, given the
large distortion of the O octahedra, however,in triangular layers
several factors of different signs contribute to $\Delta,$ and strong
cancellations are not uncommon\cite{Devina}. Trigonal splitting,
combined with various NN and NNN hoppings not accounted for above, all
of them on the order of 20 meV, trigger subband dispersions of
200--300 meV (see SI for further discussion).

We shall now address the SO interaction. The corresponding bands and
density of states (DOS) are shown in Fig.~\ref{bandSO}.  The lowest
two subbands hardly exhibit any SO effect, even though the spin-orbit
parameter $\lambda$ in Ir is $\sim$ 0.4-0.5 eV, larger than both the
subband widths and subband separation. However, a simple calculation
shows that not only are the orbital momentum matrix elements between
the QMOs on the same hexagon zero (this follows from the quenching of
the orbital momentum in the QMO states), but they also vanish between
the like QMOs, located at the neighboring hexagons, such as
$B_{1u}-B_{1u}.$ Furthermore, at $\Gamma$ the matrix elements between
the two lowest subbands, $B_{1u}$ and $E_{1g},$ vanish because of
different parities; away from the $\Gamma$ point the effect of SO
increases, in the first approximation, as $F(\mathbf{k}) =
\sin^{2}\mathbf{kA}+\sin^{2}\mathbf{kB}+\sin^{2}\mathbf{kC}$, where
\textbf{A}, \textbf{B} and \textbf{C} are the three vectors connecting
the centers of the hexagons, as can be worked out by applying the
${\bf L}\cdot{\bf S}$ operator to the corresponding QMOs.

The situation becomes more complex in the upper manifold, where three
bands, $A_{1g}$ and two $E_{2u}$, come very close. Even though the
diagonal matrix elements, as well as nondiagonal elements at $\Gamma$
still vanish, the fact that $A_{1g}$ and $E_{2u}$ are nearly
degenerate in energy induces a considerable SO effect at all other
\textbf{k}-points (which grows linearly with $k,$ as
$\sqrt{F(\mathbf{k)}}).$ Note that deviations from the minimal model
($t_{1}^{\prime},t_{2}^{\prime}$) and SO coupling with the lower
$E_{1g}$ states also affect the bands at $k=0$.  We also remind that
the orbital moment of the individual electronic states can only be
finite if the QMOs mix (which is the case), and the direction of the
orbital moment is different in different parts of the Brillouin zone:
along one of the three cardinal in-plane directions it is parallel to
the cubic $x$, along another to $y$, $etc$. Since the spin moment
tends to be parallel to the orbital moment, SO is competing with the
Hund's rule coupling and suppresses the tendency to magnetism.

Let us now discuss the effect of the Hubbard correlations. It was
initially conjectured that {\na} was a Mott insulator. This seems
counterintuitive, since similar $4d$ Ru and Rh compounds are
correlated metals, and more diffuse 5$d$ orbitals have a smaller
Hubbard $U\sim1.5-2$ eV and stronger hybridization. It is hard to
justify that this $U$ can drive a 5/6 filled band of a similar width
into an insulating state. Recently another, more logical concept has
gained currency: on the LDA level {\na} is a semimetal, barely missing
being a semiconductor, and a small Hubbard $U$ just helps to enhance
the already (spin-orbit driven) existing gap. Indeed, in our
calculations the minimal gap is $-$8 meV, but the average direct gap
is 150 meV, consistent with the optical absorption\cite{Dirk}. The
minimal direct (optical) gap is 50 meV, so it is plausible that it is
somewhat enhanced by correlation effects.

In order to include the effect of an onsite Hubbard $U$ in the QMO
basis, a $U_{\rm QMO}\sim U/6$ has to be applied to each
QMO\cite{Gunnarsson}, with a residual Coulomb repulsion between
neighboring QMOs, $V_{\rm QMO}\sim U/18=U_{\rm QMO}/3$ (note that two
QMOs overlap on two sites). Overall, we expect that the effect of the
Coulomb repulsion in our system is similar to that in a single-site
two-orbital Hubbard model at half filling (the upper QMO band is
half-filled) and $U_{\rm QMO}\approx W\approx150-200$ meV. In this
case, since $U_{\rm QMO}$ does not compete with one-electron hopping
any more, one should expect that the gap will be enhanced by a
considerable fraction of $U_{\rm QMO}$, which is consistent with the
experiment. Thus, Hubbard correlations are of no qualitative
importance, and only moderately enhance the existing gap.

Since all electrons are fully delocalized over six sites, any model
assuming localized spins (whether Heisenberg or Kitaev) is difficult
to justify. On the other hand, the QMOs are not magnetically rigid
objects and neighboring QMOs overlap on 2 out of 6 sites, which makes
a model with magnetic moments localized on QMOs equally
unsuitable~\cite{Comment_Fe}.

\begin{figure}[tbh]
\begin{center}
\includegraphics[angle=-90,width=0.95\columnwidth]{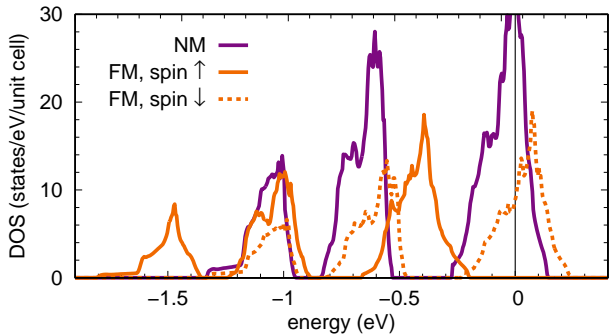}
\end{center}
\caption{Non-relativistic non-magnetic (purple) and ferromagnetic
  (orange) density of states (DOS) of {\na} calculated with the FPLO
  basis.}
\label{DOS}
\end{figure}

We will consider therefore magnetism in the itinerant approach.  In
the non-relativistic case, the non-magnetic DOS shows a high peak at
$E_F$ due to $E_{2u}$ and $A_{1g}$ merging and rather flat band
dispersion (see Fig.~\ref{bandSO}). Such a system is unstable against
ferromagnetism (FM) and the peak is easily split gaining exchange
energy (1 $\mu_{\rm B}/$Ir) with little loss of kinetic energy.  The
resulting FM state is half-metallic (Fig.~\ref{DOS}) (see SI).

Turning on the SO interaction has a drastic effect on magnetism.  SO
competes with the Hund's rule that favors all onsite orbitals to be
collinear. The spin moment is then reduced from $1$ $\mu_{\rm B}$ to
$\approx0.4$ $\mu_{\rm B}$/Ir for ferro-, and $\approx0.2$ $\mu_{\rm
  B}$/Ir for the zigzag and stripe antiferromagnetic (AFM)
arrangements (see SI). The orbital moment is parallel to the spin one,
reminiscent of the $j_{\rm eff}=1/2$ state, and is roughly equal in
magnitude and not twice larger, as it should be for $j_{\rm eff}=1/2$.
The energy gain for the FM case drops to a few meV/Ir~\cite{gain}, and
the zigzag pattern evolves as the most favorable AFM state.

Qualitatively, two closely competing ground states emerge from the
relativistic DFT calculations: ferromagnetic and zigzag.  In the
context of an itinerant picture, we can argue as follows. SO creates a
pseudogap at the Fermi level in the non-magnetic calculations (see
Fig.~\ref{bandSO}).  This gains one-electron energy and any AFM
arrangement that destroys this pseudogap incurs a penalty.  From the
three considered AFM states only zigzag preserves (even slightly
enhances) the pseudogap (see SI).  That gives this state an immediate
energetical advantage and leads to the energy balance described
above. Two notes are in place: first, all the above holds in LDA+U
calculations with a reasonable atomic $U$ (we have checked $U$ up to
3.8 eV). The role of $U$ in these systems - as stated previously- is
merely enhancing the existing SO-driven gap.  Second, if the DOS
indeed plays a decisive role in magnetic interactions, it is unlikely
that they can be meaningfully mapped onto a short-range exchange
model, Heisenberg or otherwise.

Summarizing, our DFT calculations demonstrate that {\na} is close to
an itinerant regime. The electronic structure of this system is
naturally described on the basis of quasi-molecular orbitals centered
each on its own hexagon. This makes this, and similar materials rather
unique. Proceeding from this description one can understand the main
properties of {\na}, including its unique zigzag magnetic ordering
with small magnetic moment.

However, the main goal of our work is not a complete understanding of
the magnetic properties of {\na}. We realize that this understanding
is still incomplete and that full explanation of the weak
antiferromagnetism, as well as of the magnetic response in this
compound remains a challenge. Rather, we lay out the framework in
which this challenge has to be resolved. We demonstrate that both the
simplified (but correct) TB model proposed in previous
studies~\cite{Jackeli2009,Shitade2009}, and full {\it ab initio}
calculations provide a framework that is best described by the
quasi-molecular orbitals. This is an as yet unexplored concept (as
opposed to molecular orbitals or atomic orbitals), and there are many
open questions about how to treat correlations, magnetic response {\it
  etc.} in this framework, however, it appears to be the only way to
reduce the full 12 atomic orbitals ($t_{2g}$ or their relativitsic
combinations) problem to a smaller subspace (3$\times 2=6$) QMOs.

I.I.M. acknowledges many stimulating discussions with Radu Coldea and
his group, and with Alexey Kolmogorov, and is particularly thankful to
Radu Coldea for introducing him to the world of quasihexagonal
iridates. H.O.J., R.V. and D.Kh. acknowledge support by the Deutsche
Forschungsgemeinschaft through grants SFB/TR 49 and FOR 1346
(H.O.J. and R.V.) and SFB 608 and FOR 1346
(D.Kh.). H.O.J. acknowledges support by the Helmholtz Association via
HA216/EMMI.









\newpage

\section{Supplementary Information}
\renewcommand{\thefigure}{S\arabic{figure}}
\setcounter{figure}{0}
\renewcommand{\thetable}{S\Roman{table}}
\setcounter{table}{0}

\begin{figure}[tbh]
\caption{ Crystal structure of {\na} in the cubic setting. The
hexagonal direction is along the [111] direction in this setting. Ir, O and Na
atoms are shown as grey, magenta, and yellow spheres, respectively. The
three inequivalent Ir-Ir bonds are labeled according to their cubic
directions.} 
\label{structure}
\begin{center}
\includegraphics[width=0.9 \columnwidth]{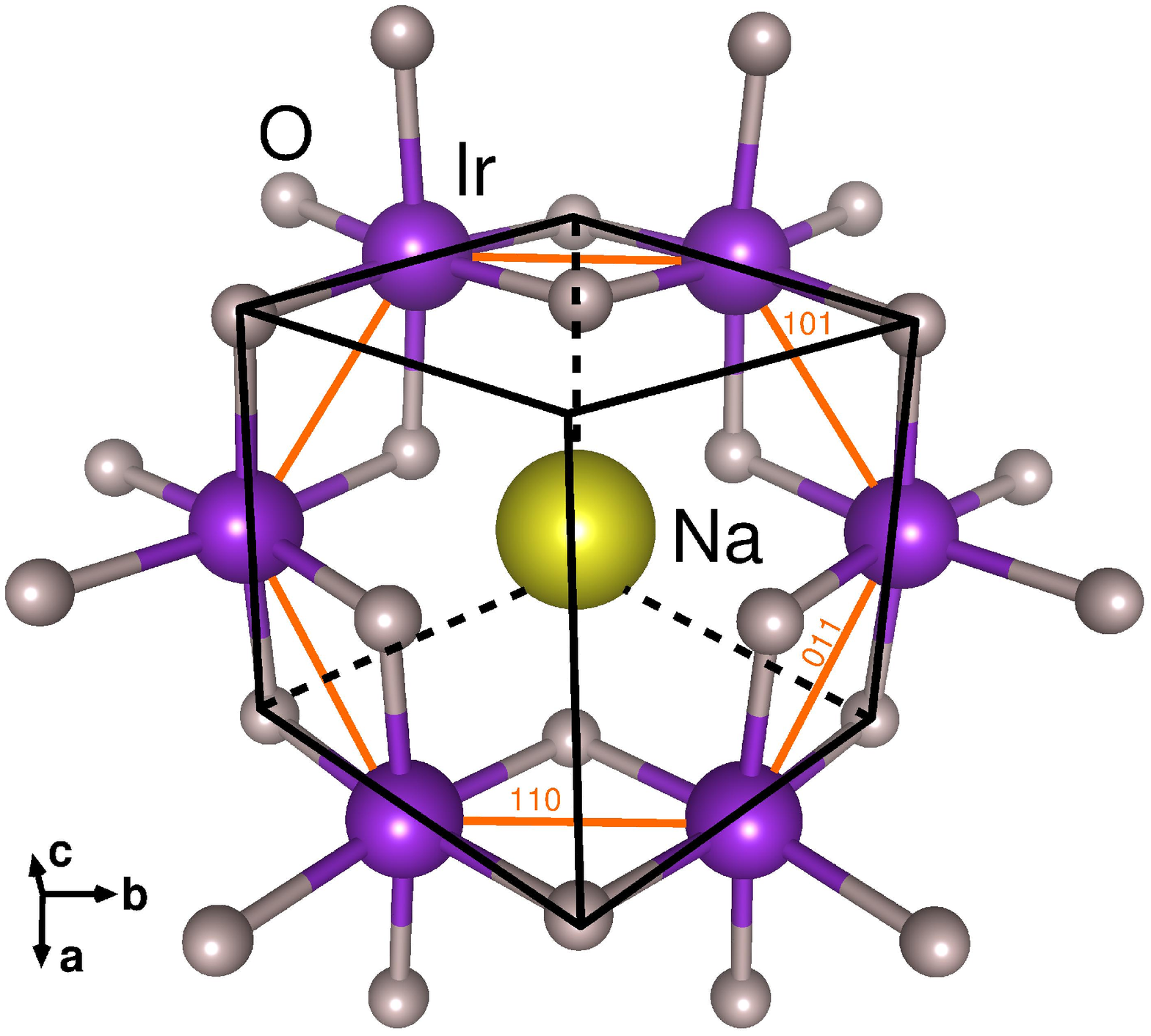} %
\end{center}
\end{figure}

We performed density functional theory (DFT) calculations considering
various full potential all electron codes, such as
WIEN2k~\cite{wien2k}, ELK~\cite{elk}, and FPLO~\cite{fplo} using the
generalized gradient approximation functional in its PBE
form~\cite{Perdew96}, and verified that the results agree reasonably
well among different codes. Such comparison is particularly important
because the codes implement the spin-orbit coupling in slightly
different ways, employing usually unimportant, but in principle
unequal approximations. In the \textit{\ non-relativistic}
calculations the core electrons were treated fully relativistically
and the valence electrons non-relativistically (scalar relativistic
approximation). In the \textit{fully relativistic} calculations,
\textit{i.e.} with inclusion of spin-orbit coupling, all electrons
were treated fully relativistically.  We considered the C$2/m$ crystal
structure as given in Ref.~\onlinecite{Choi2011} and shown in
Fig.~\ref{structure}.

\begin{figure}[bt]
\includegraphics[angle=-90,width=\columnwidth]{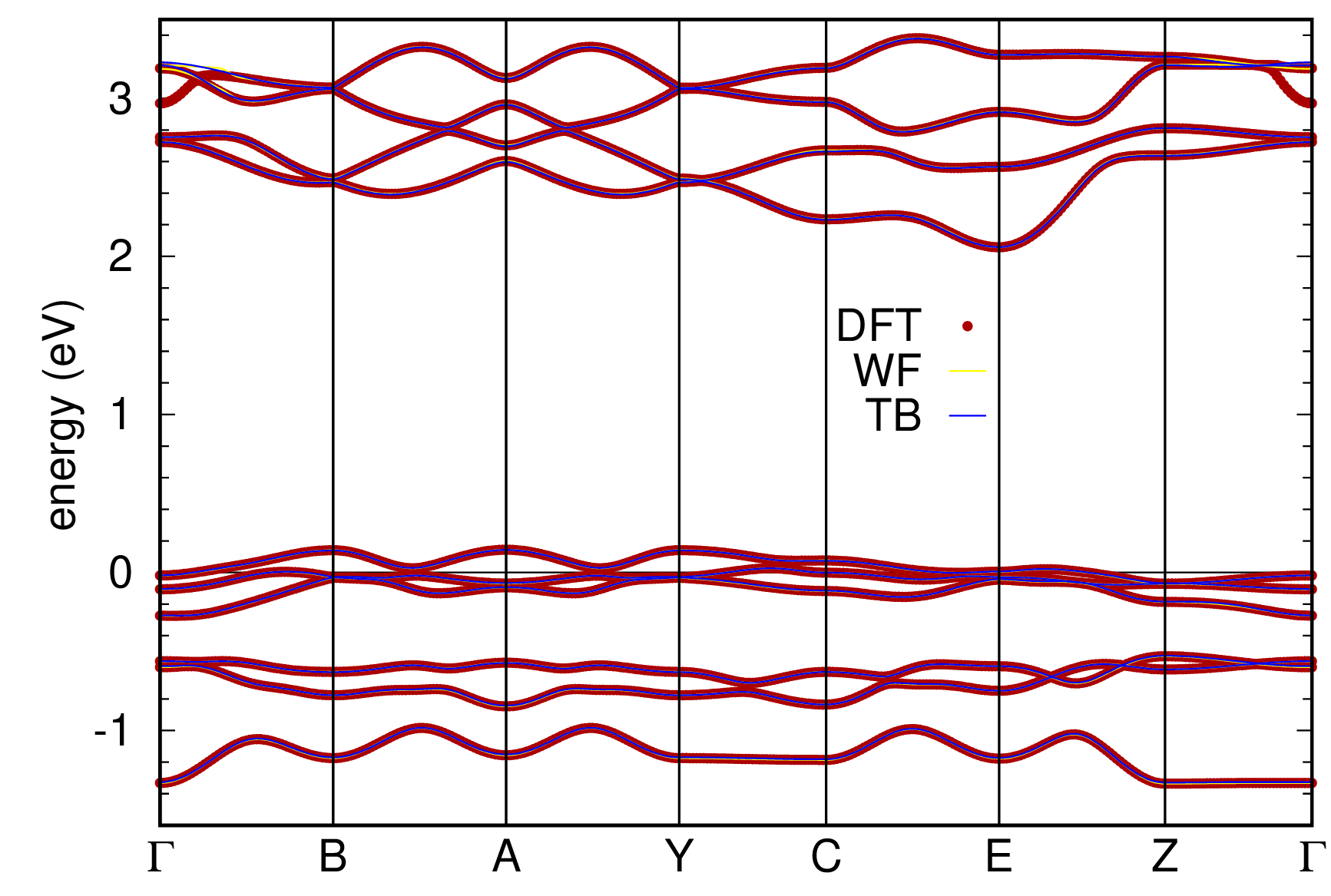}
\caption{Non-relativistic non-magnetic band structure of {\na} (red symbols)  
shown together with the Wannier bands (yellow) and the tight-binding bands (blue). }\label{fig:wannierbs}
\end{figure}

We used projective Wannier functions as implemented in the FPLO
basis~\cite{Eschrig2009} to determine a tight-binding (TB)
representation for the Ir $5d$ bands.  In Figure~\ref{fig:wannierbs}
we show the DFT band structure together with the bands corresponding
to the Wannier representation and the TB bands derived from this
representation.

\begin{figure*}[htb]
\includegraphics[width=0.3\textwidth]{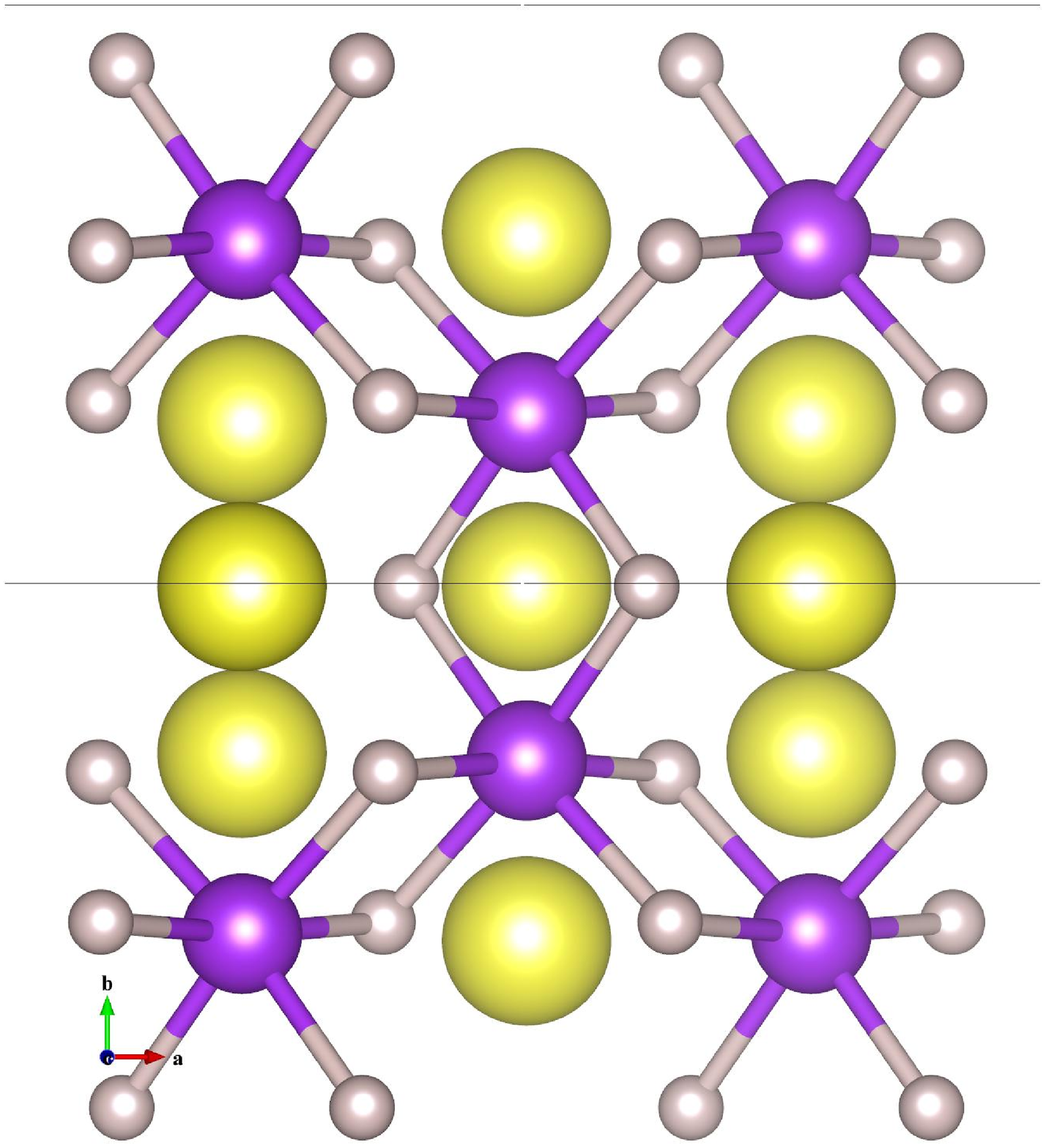}
\includegraphics[width=0.3\textwidth]{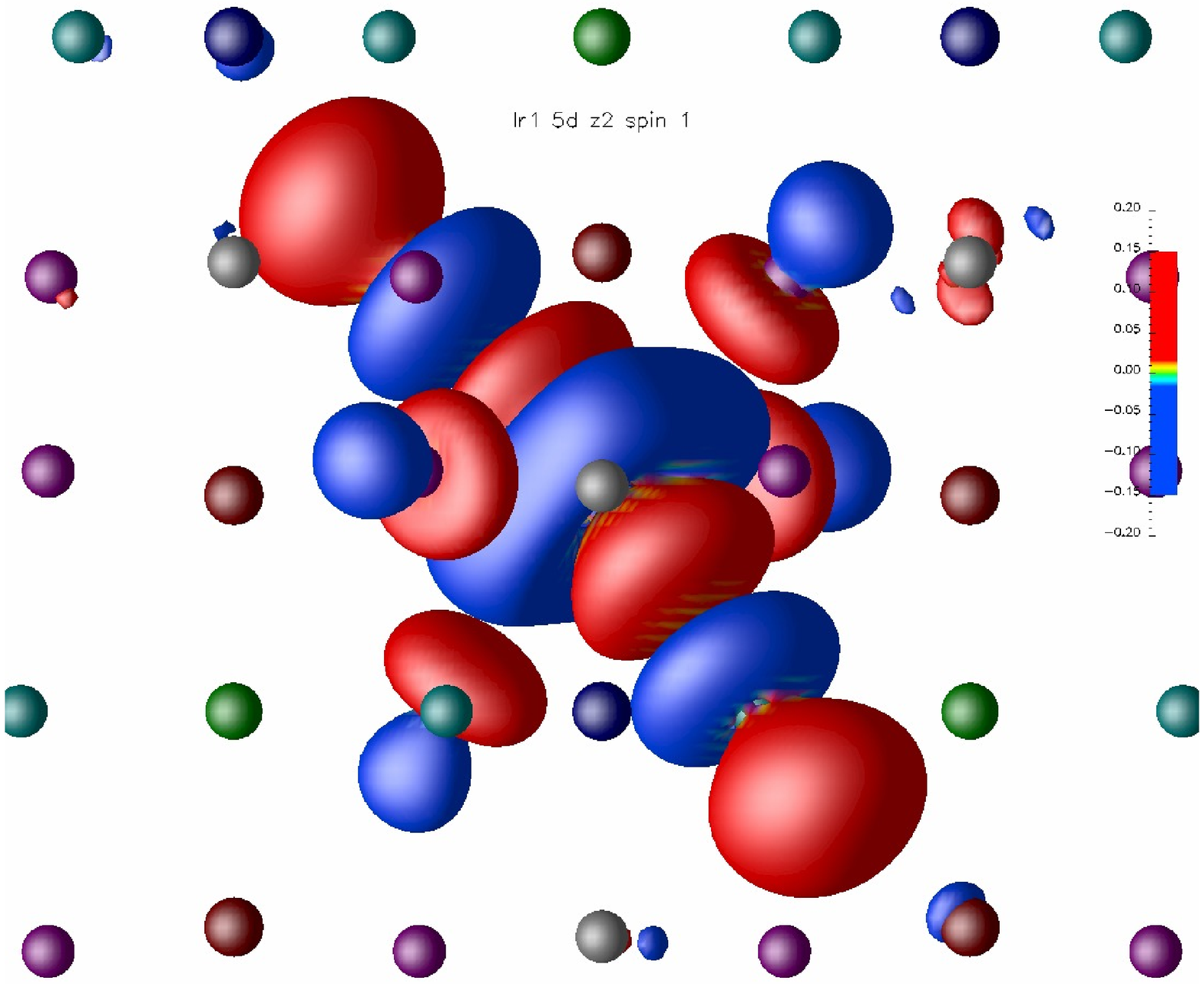}
\includegraphics[width=0.3\textwidth]{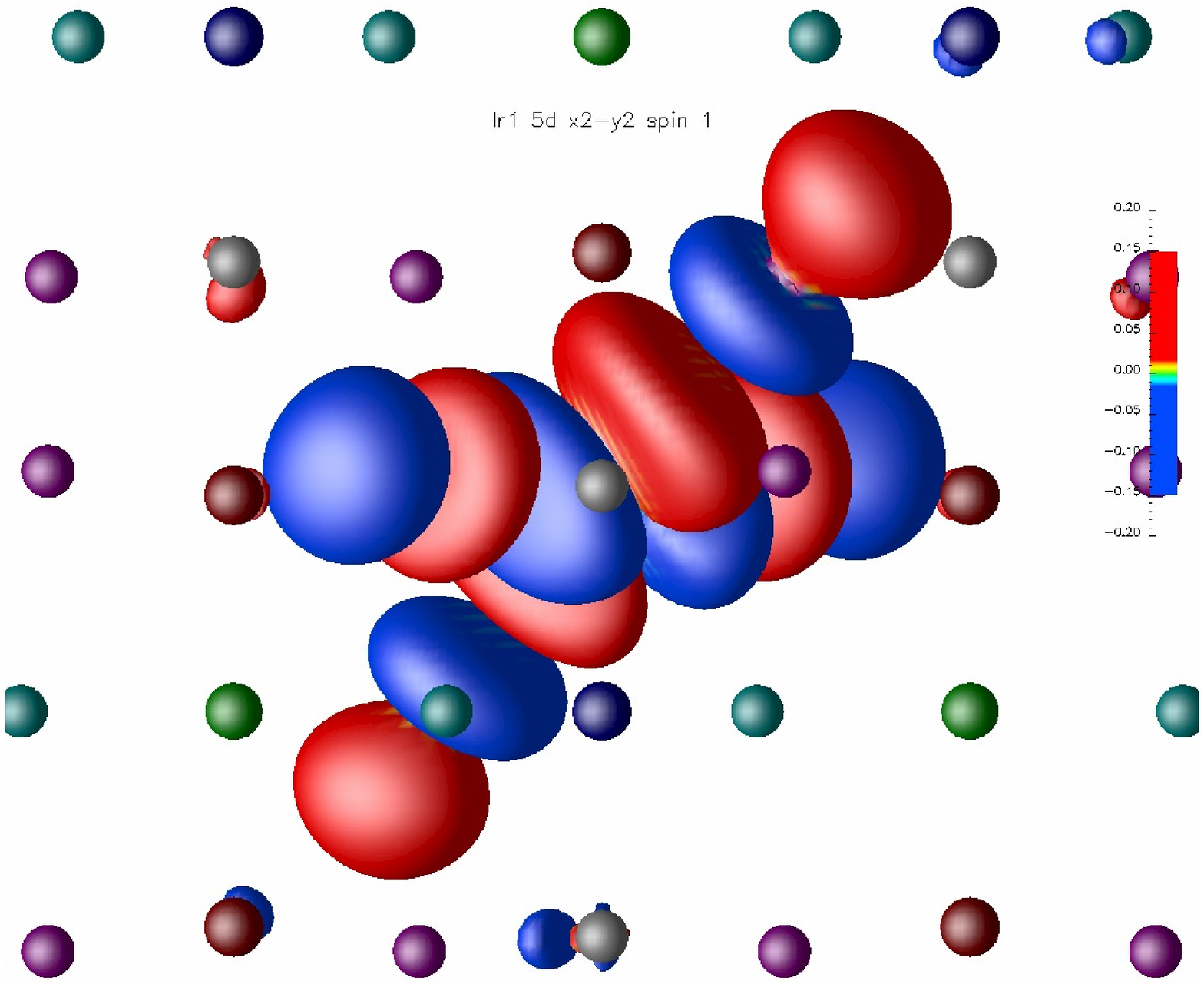}\\
\includegraphics[width=0.3\textwidth]{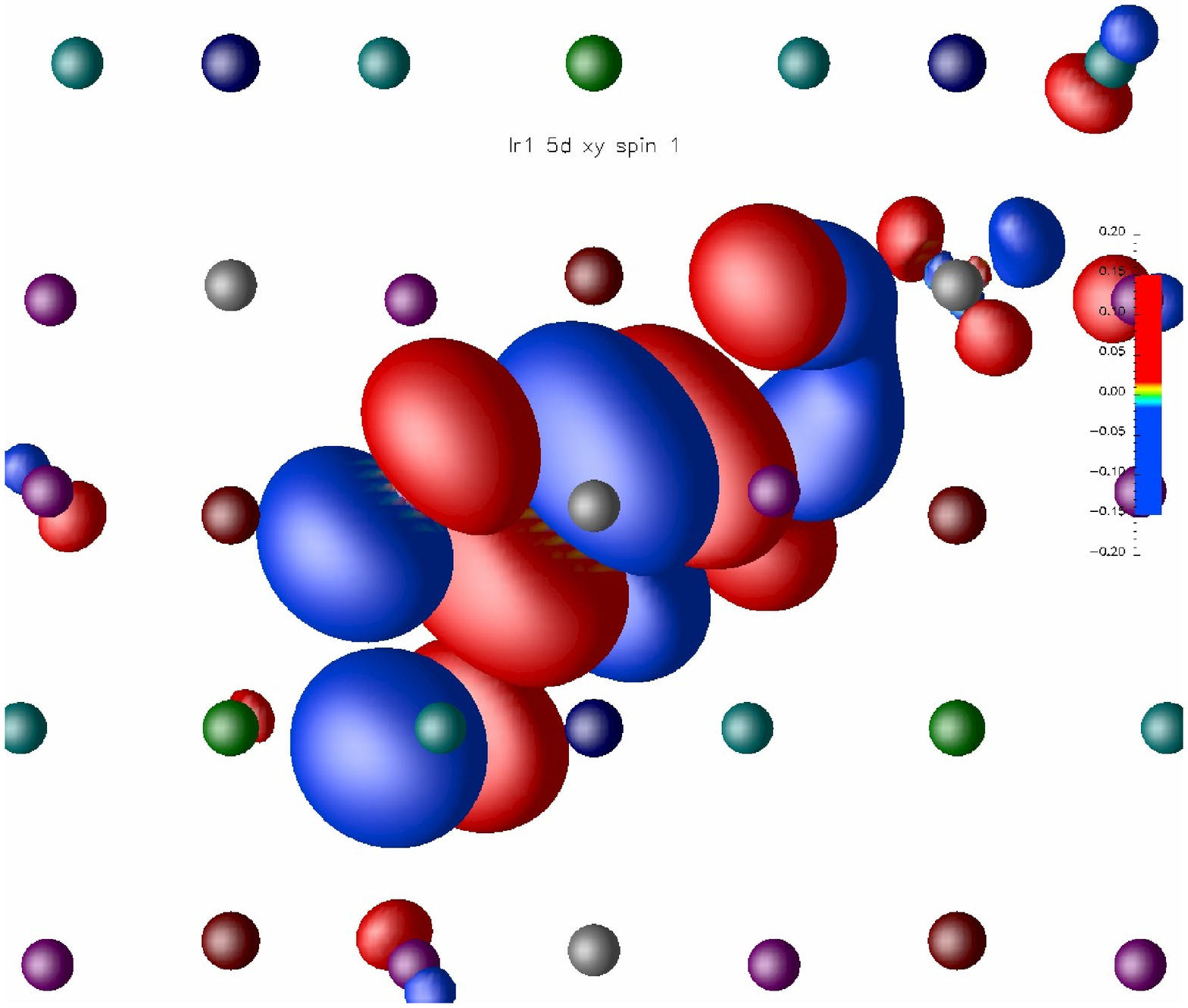}
\includegraphics[width=0.3\textwidth]{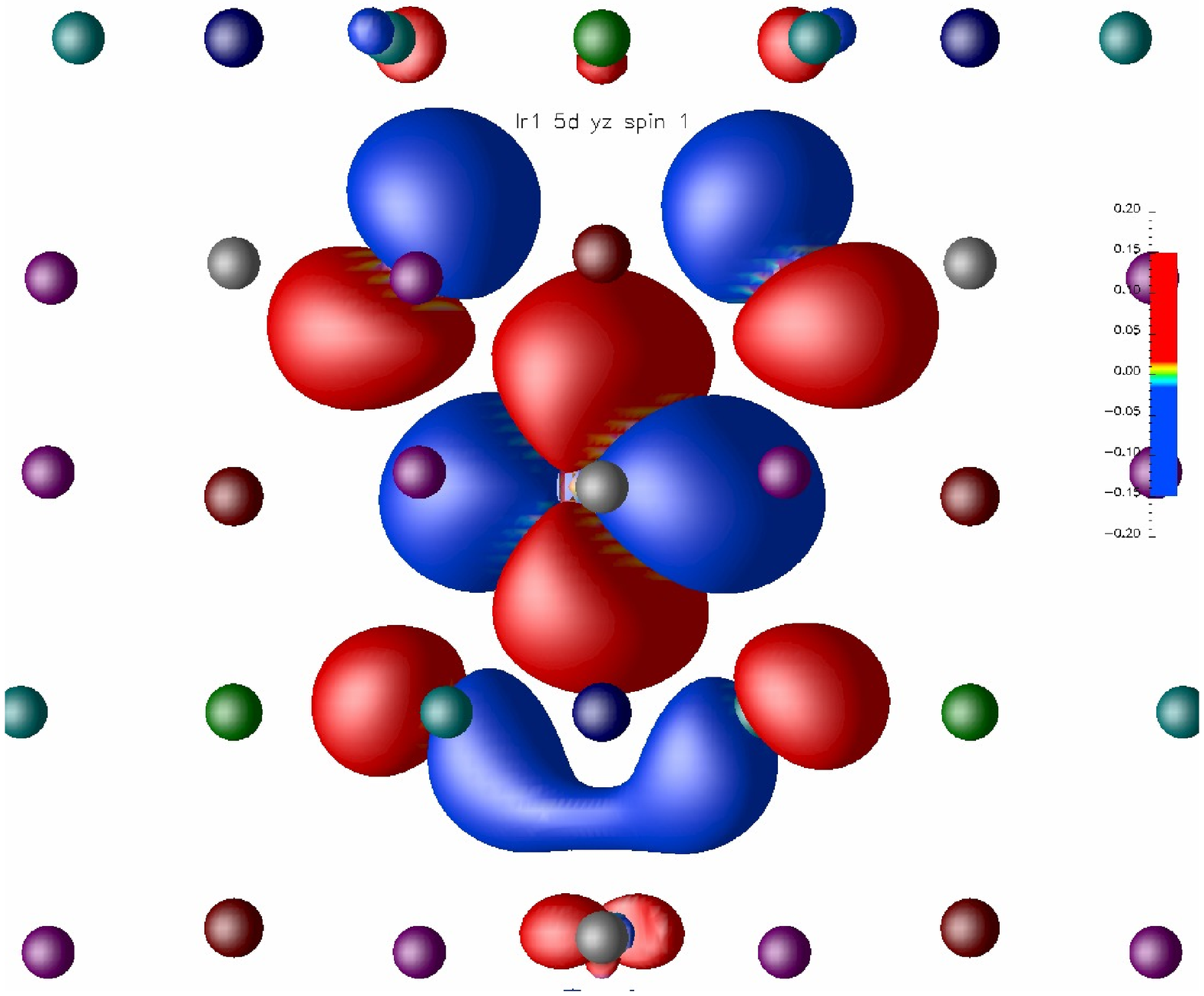}
\includegraphics[width=0.3\textwidth]{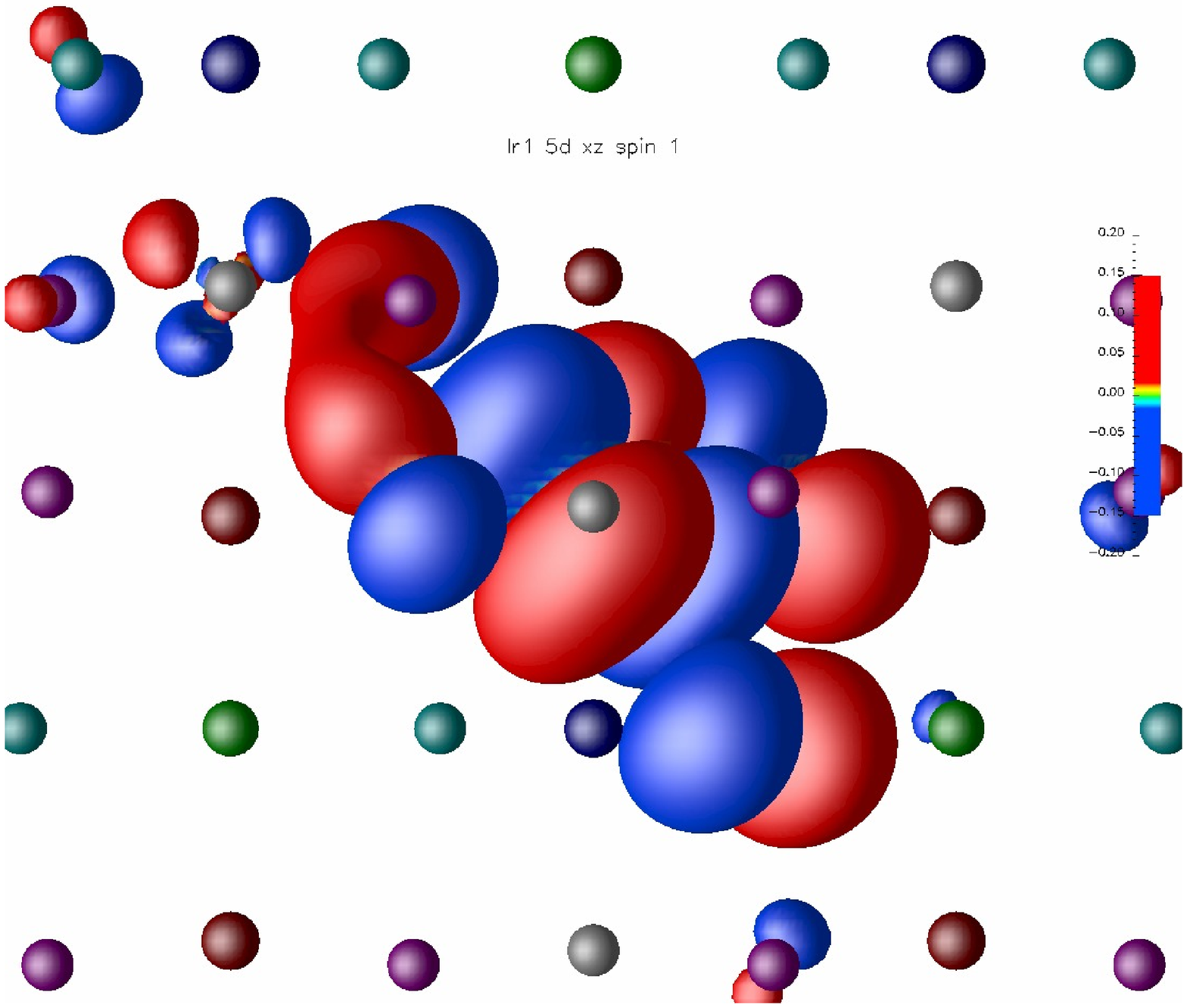}
\caption{Projective Wannier functions for five of the ten Ir $5d$ bands, together with a structure showing the perspective.}\label{fig:wannier}
\end{figure*}

In Fig.~\ref{fig:wannier} we present the projective Wannier functions
for the $5d$ orbitals of one Ir site.  The Wannier functions exhibit
the typical shape of the $5d$ functions at the Ir site. Besides, they
show a clear asymmetry due to Na as well as tails on the O sites.

\begin{figure*}[htb]
\includegraphics[angle=-90,width=0.5\textwidth]{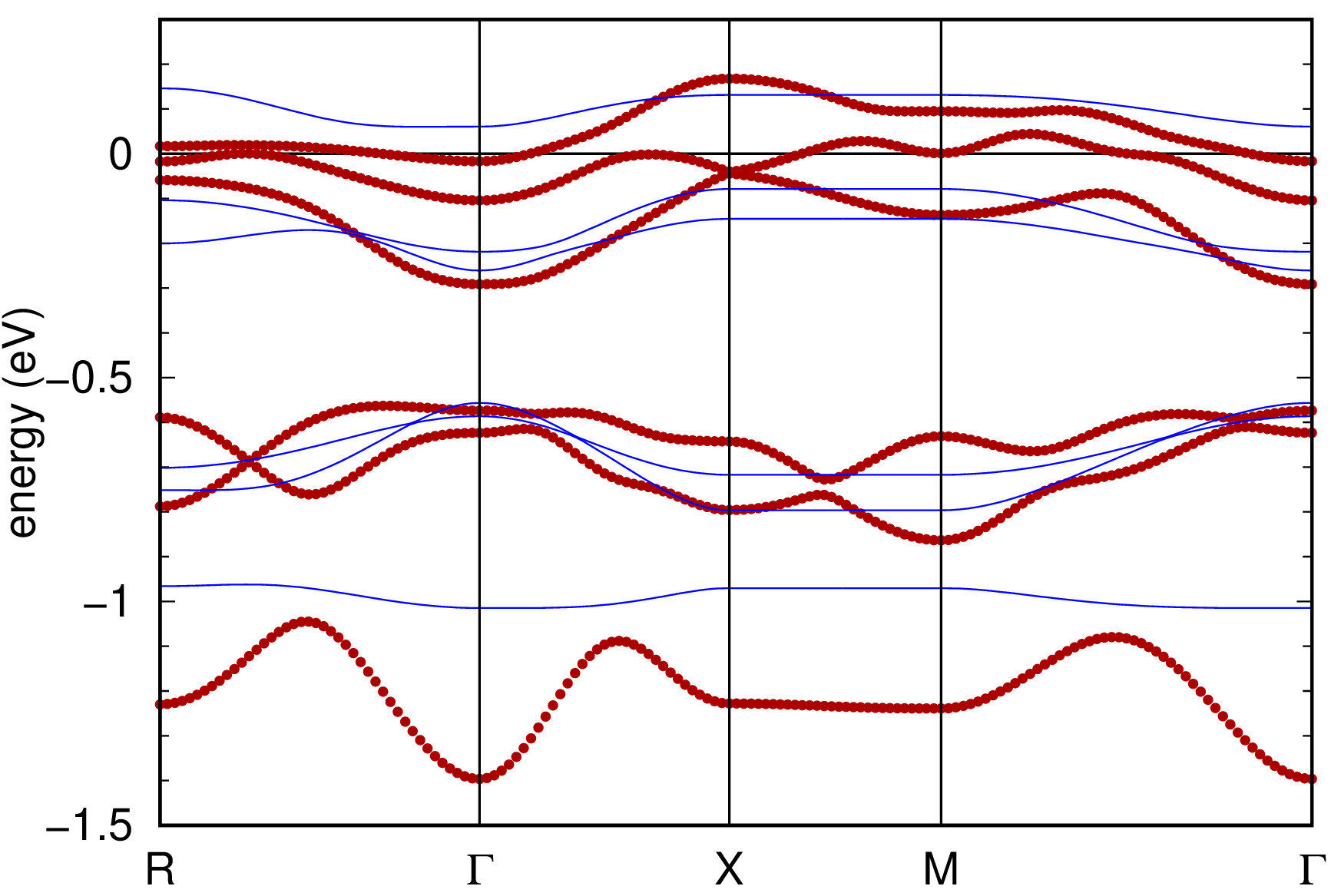}\includegraphics[angle=-90,width=0.5\textwidth]{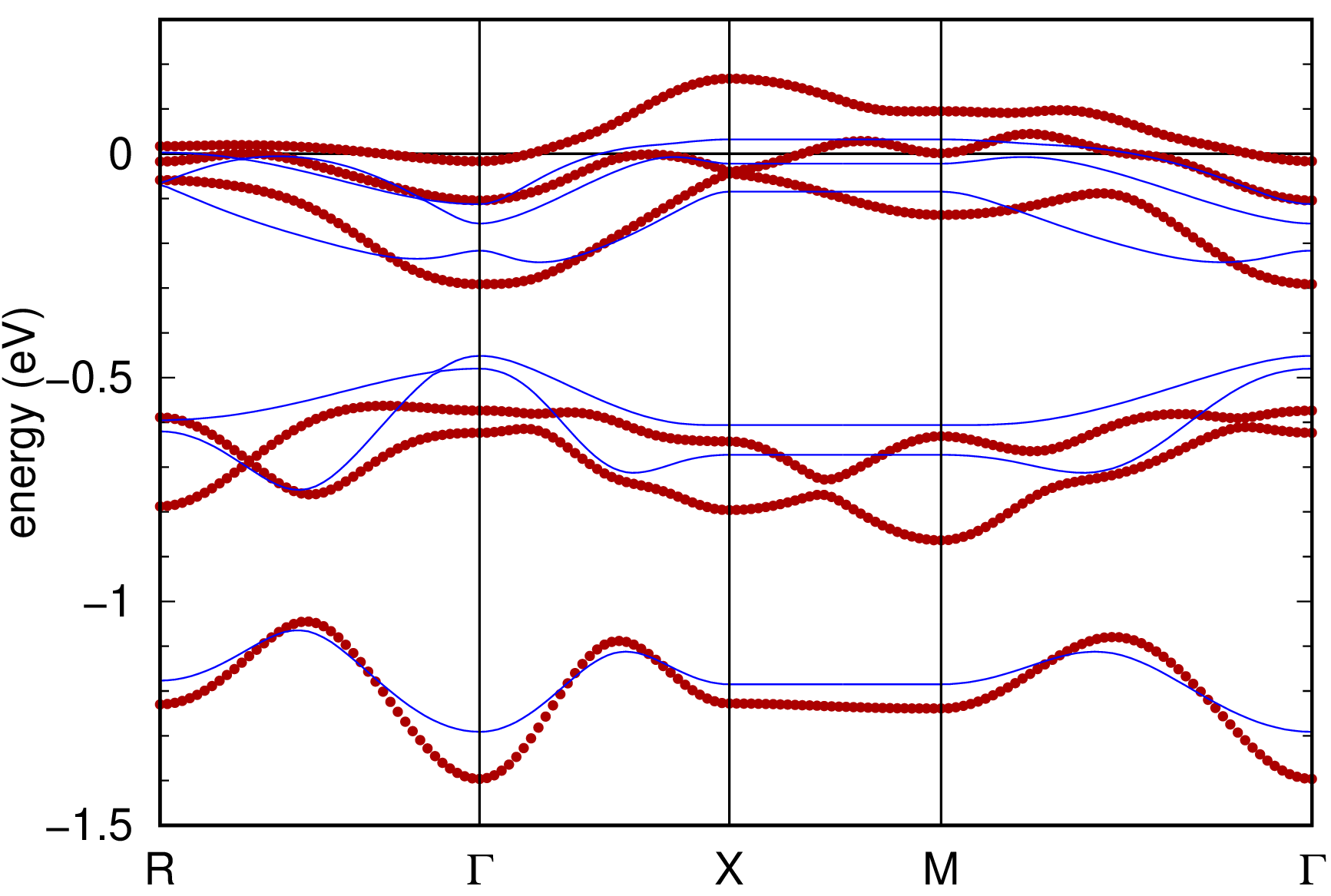}\\
\includegraphics[angle=-90,width=0.5\textwidth]{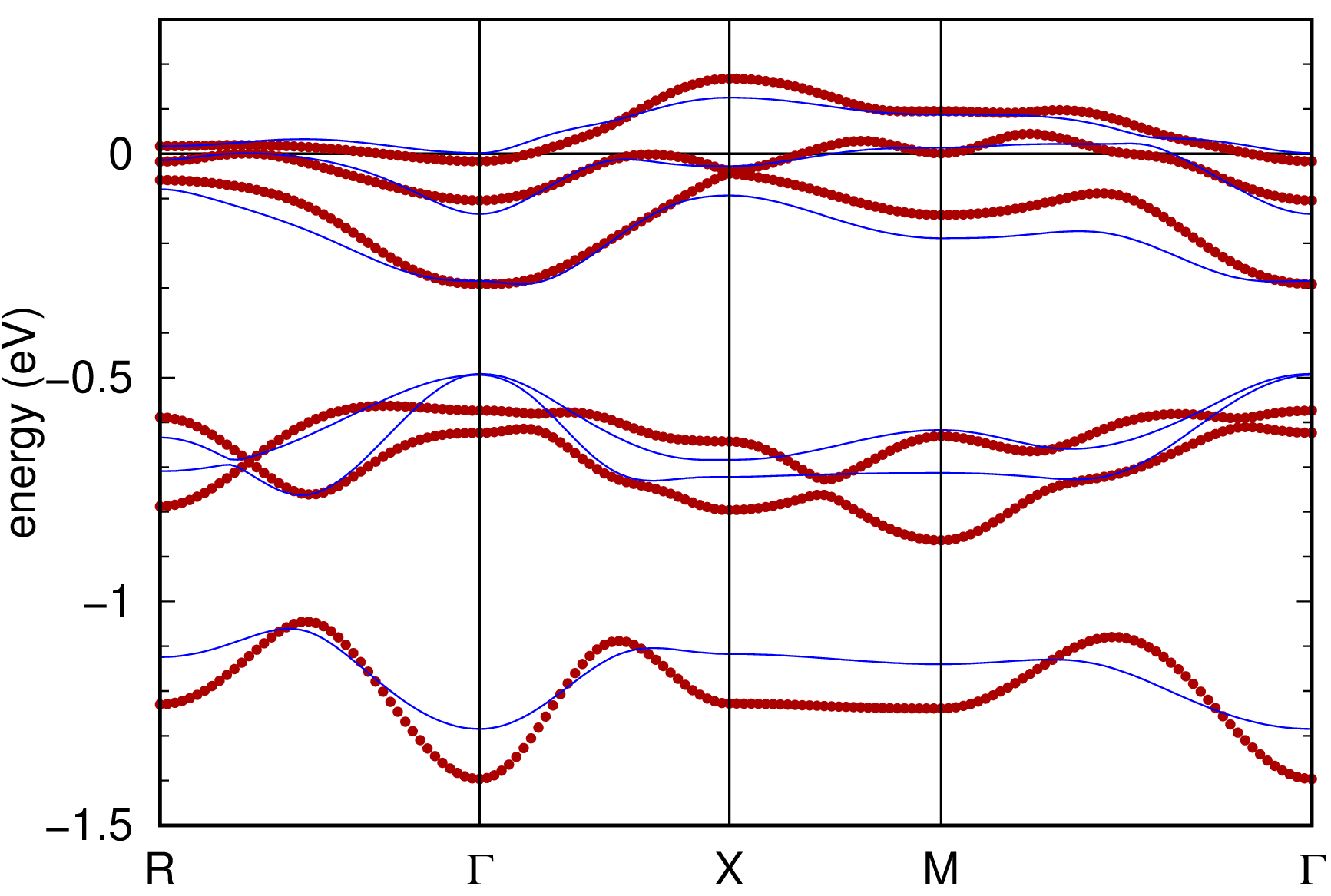}\includegraphics[angle=-90,width=0.5\textwidth]{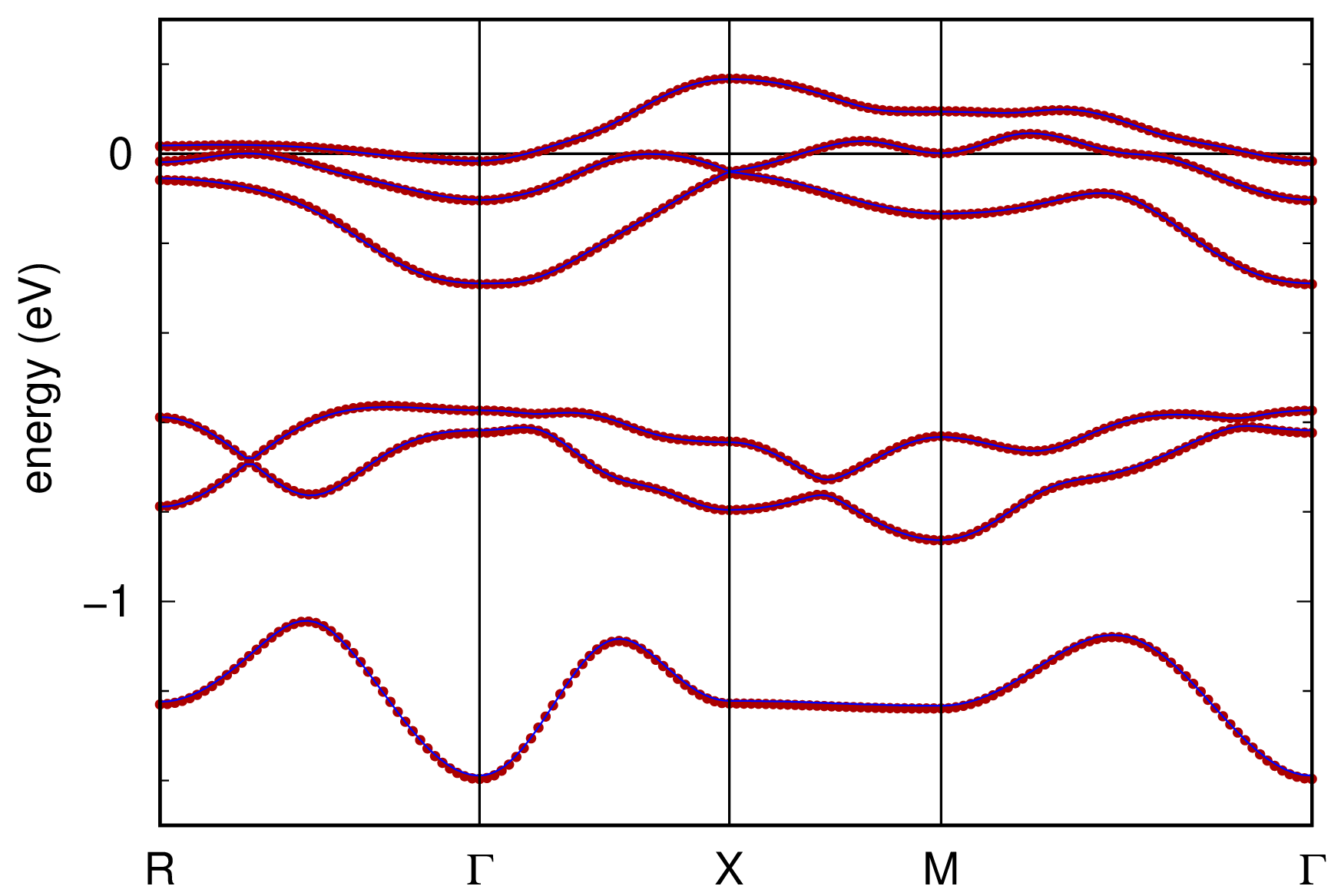}\\
\caption{Band structure of {\na} (red symbols) shown together with
 the tight-binding models that include only nearest neighbors (top left), up to next nearest neighbors (top right), up to third nearest neighbors (bottom left) and neighbors up to 16~{\AA} (bottom right). }\label{fig:restrictedbs}
\end{figure*}

In order to analyze the contribution to the non-relativistic band
structure of the various tight-binding hopping parameters and its
relation to the quasi-molecular orbital (QMO) picture, we present in
Fig.~\ref{fig:restrictedbs} the band structure that results if we
restrict the tight-binding Hamiltonian to first neighbors (top left),
up to second nearest neighbors (top right), up to third nearest
neighbors (bottom left), and without restriction (bottom right).  One
can see that already the second neighbors model provides a good
semiquantitative description of the band formation.

In the next Figure~\ref{fig:qmobs} we show the tight-binding band
structures within the QMO model. In these calculations we have
included the on-site trigonal splitting (the top left panel), adding
the nearest neighbors $t_1^\prime$ hopping (top right), then the
second nearest neighbors $t_2^\prime$ hopping (bottom left) and,
finally, including also the third nearest neighbors hopping between
the like orbital, which also proceeds through Na and does not take an
electron out of the corresponding QMO (bottom right). The small
dispersion that arises for nearest neighbors is due to deviations from
the perfect octahedral environment of iridium. Upon inclusion of
second nearest neighbors, as mentioned in the main text, the upper
doublet and singlet merge to form one three-band manifold.

\begin{figure*}[htb]
\includegraphics[angle=-90,width=0.5\textwidth]{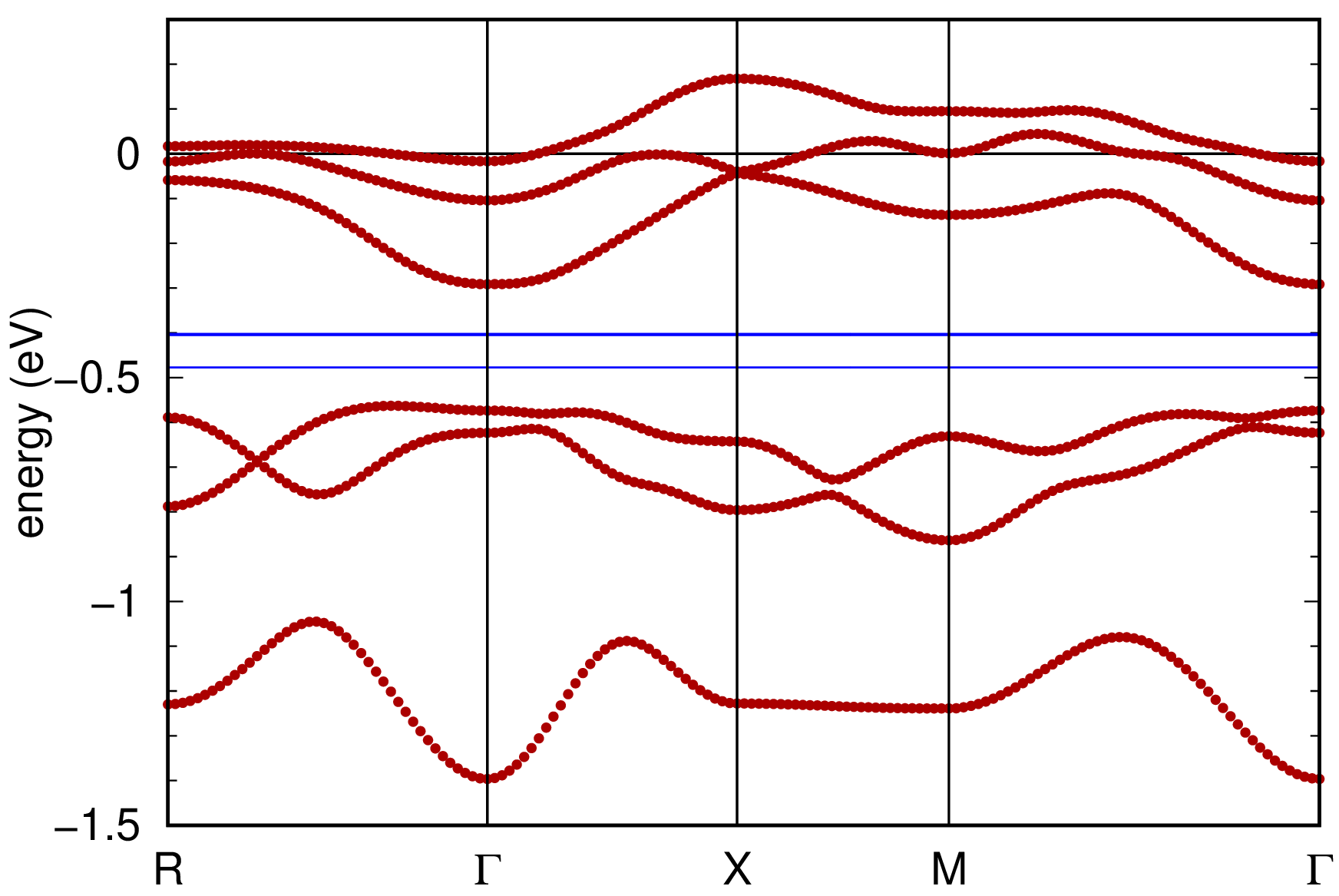}\includegraphics[angle=-90,width=0.5\textwidth]{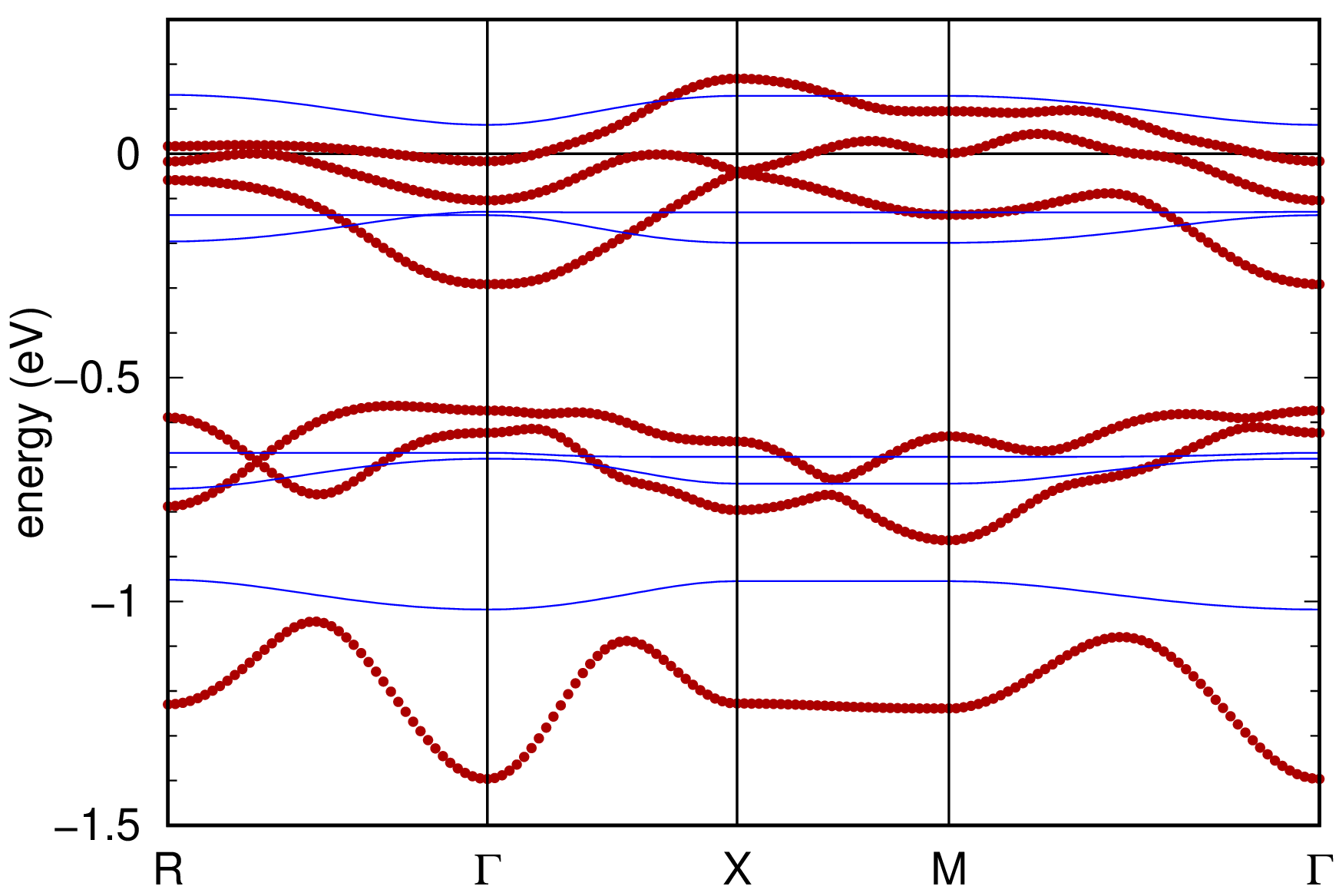}\\
\includegraphics[angle=-90,width=0.5\textwidth]{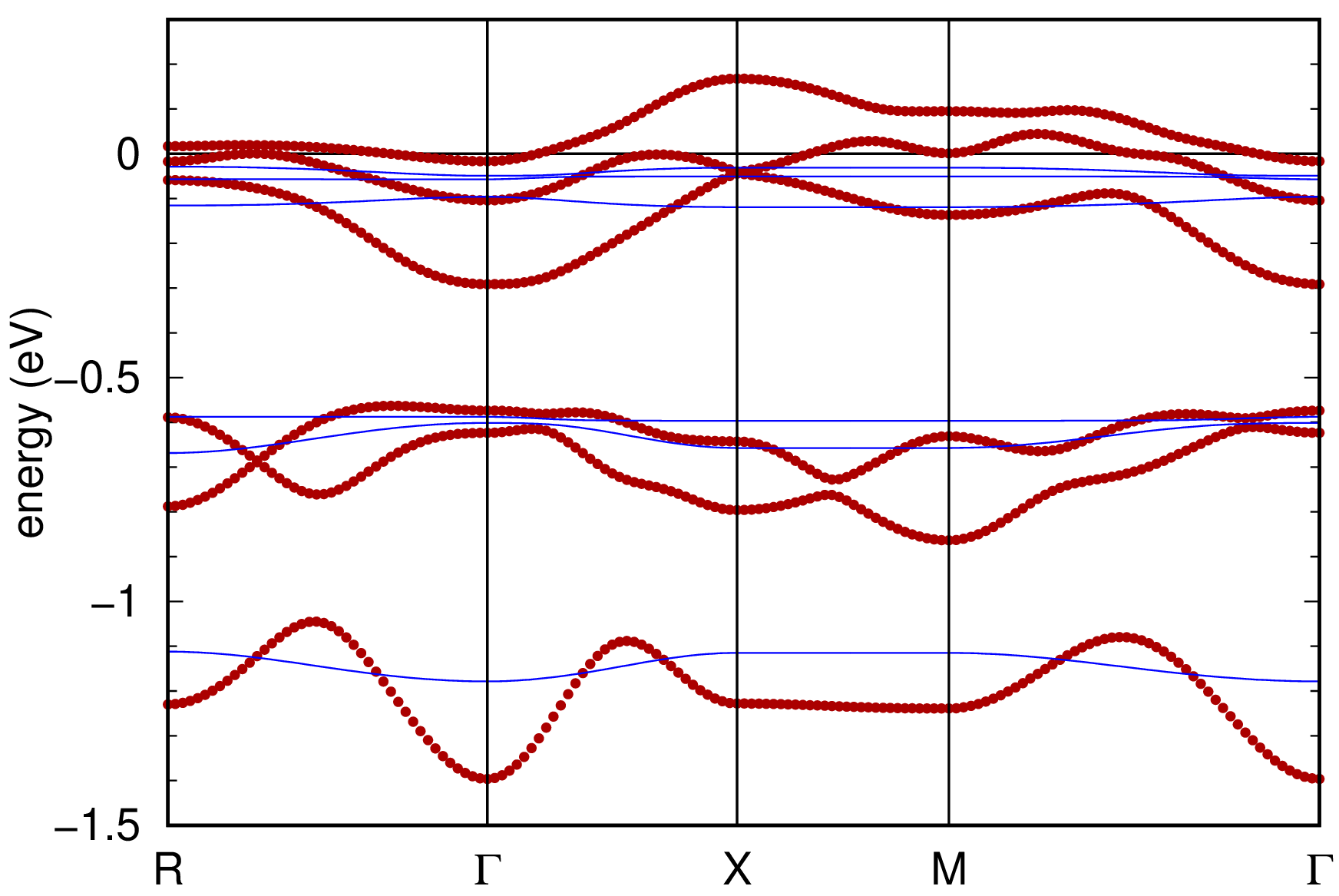}\includegraphics[angle=-90,width=0.5\textwidth]{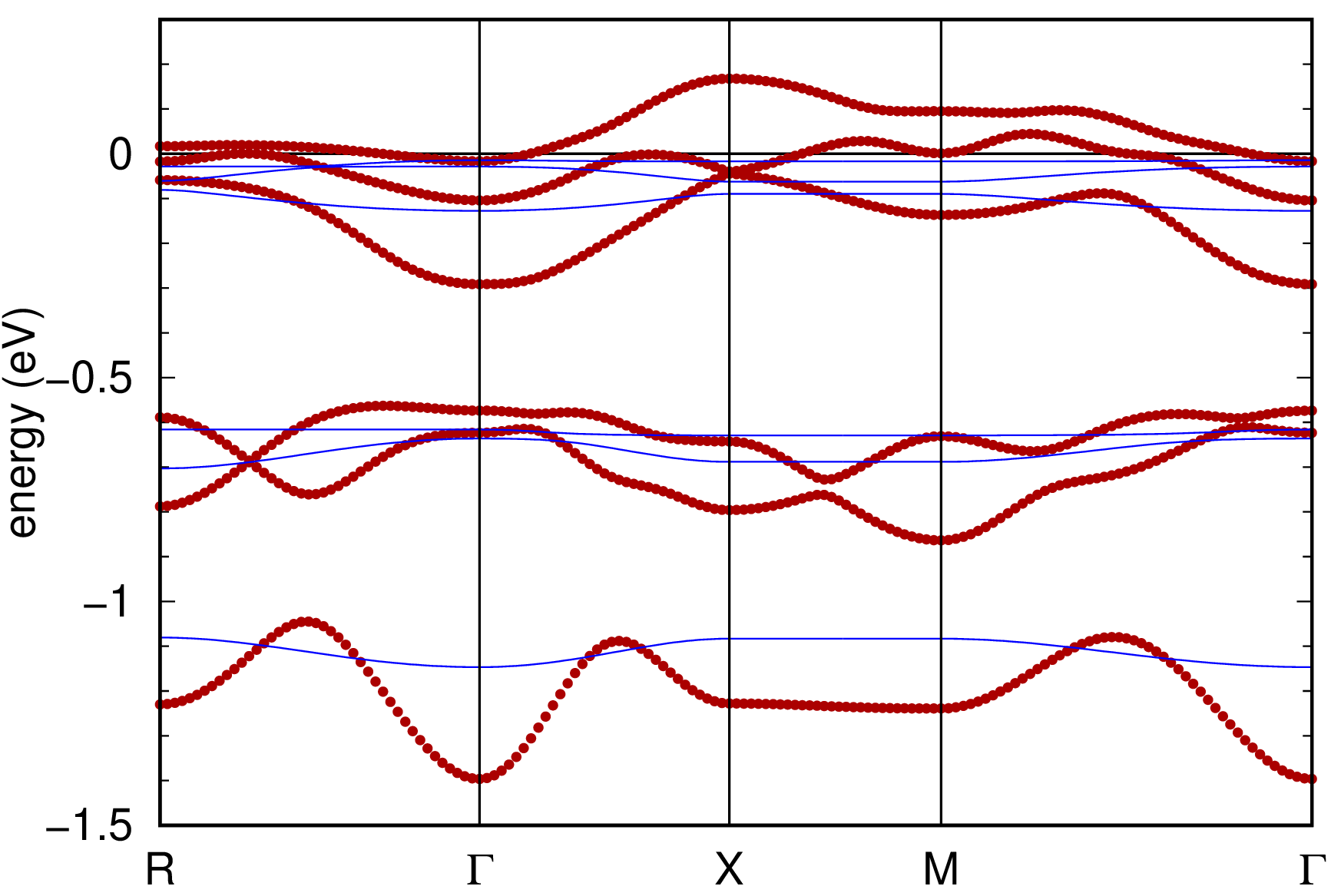}\\
\caption{Band structure of {\na} (red symbols) shown together with the
  tight-binding models that involve only parameters compatible with
  the quasi-molecular orbitals.  Only on-site parameters (top left), up
  to nearest neighbors (top right), up to second nearest neighbors
  (bottom left) and up to third nearest neighbors (bottom
  right). }\label{fig:qmobs}
\end{figure*}

\begin{figure}[tbh]
\begin{center}
\includegraphics[angle=-90,width=0.95\columnwidth]{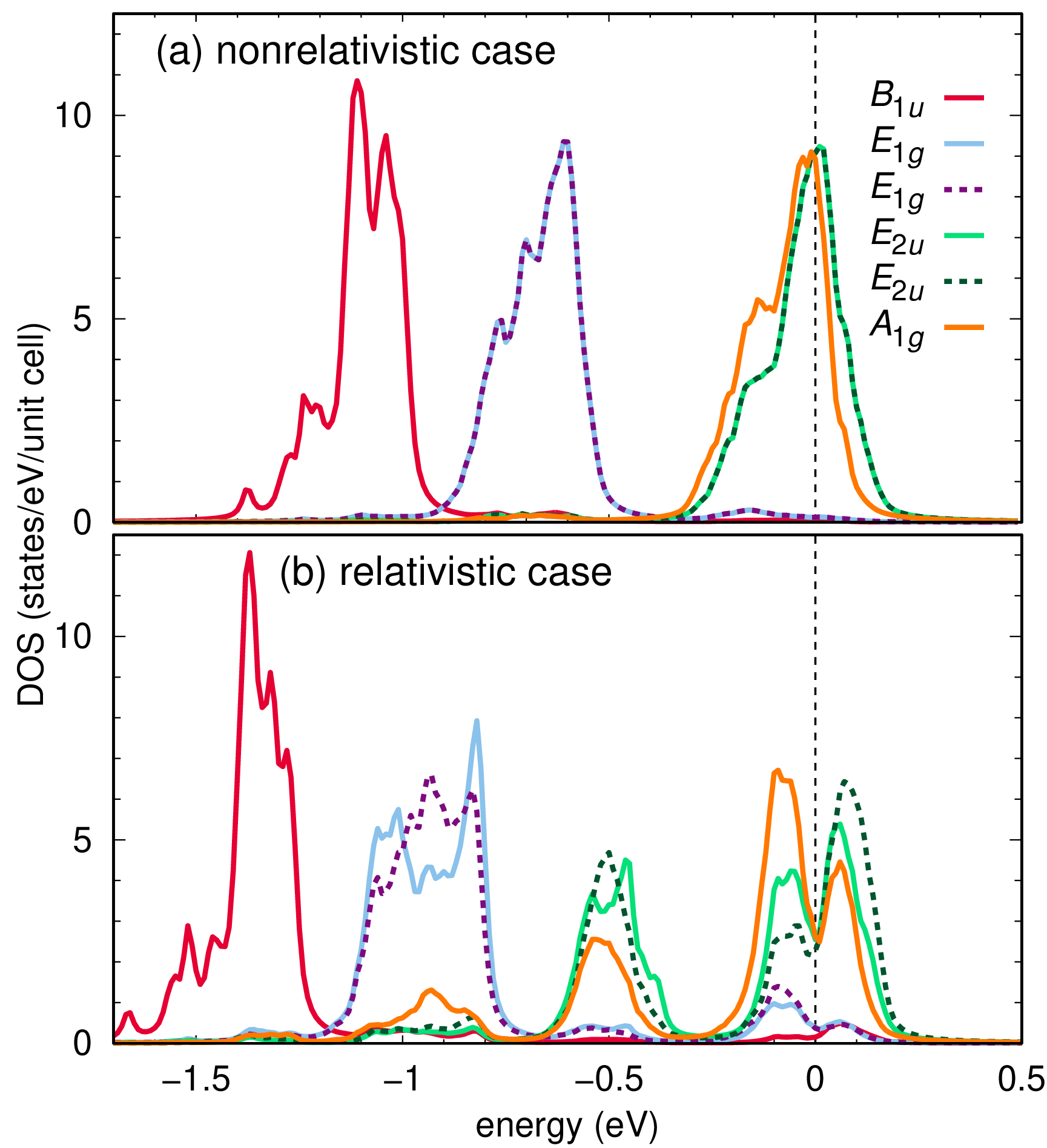}
\end{center}
\caption{Density of states of {\na} projected onto the six
  quasi-molecular orbitals given in Table~[1] of the main text 
for (a) a
  nonrelativistic and (b) a relativistic calculation.}
\label{projection}
\end{figure}

In Figure~\ref{projection}, we show projections of the total density
of states of {\na} onto the quasi-molecular orbitals specified in
Table~[1] of the main text.
The eigenvector matrix
\begin{equation*}
U=
\begin{pmatrix}
1&1&1&1&1&1\\
1& \omega& \omega^{2}&-1& \omega^{4} & \omega^{5}\\
1& \omega^{5}& \omega^{4}& -1& \omega^{2}& \omega \\
1& \omega^{2}& \omega ^{4}& 1& \omega^{2}& \omega^{4}\\
1& \omega^{4}& \omega^{2}& 1& \omega^{4}& \omega^{2}\\
1&-1&1&-1&1&-1
\end{pmatrix}
\end{equation*}
(with $\omega=\exp(i\pi/3)$) is a unitary transformation that rotates
the atomic Ir $t_{2g}$ orbitals into the QMO orbital space. 
$E_{1g}$
and $E_{2u}$ states are perfectly degenerate in the nonrelativistic
case (Figure~\ref{projection} (a)). 
When spin-orbit coupling is turned on (Figure~\ref{projection} (b)), interestingly, the three upper bands are
no more equivalent in this sense, with the central band being mostly $A_{1g}$, and the other 
two mostly $E_{2u}$. Importantly, there is hardly any mixing between the lower
three bands and the upper three bands, emphasizing the fact that the low-energy
physics is nearly exclusively defined by the upper three QMOs, and their mutual interaction, 
whether with or without spin-orbit. 
At the same time, one can, alternatively, project the same bands
onto the relativistic orbitals, $j_{eff}=1/2$ and $j_{eff}=3/2$, and, as observed
before\cite{kunes}, the upper two bands have more  $j_{eff}=1/2$ character than
 $j_{eff}=3/2$ character, but, for instance, at the $Gamma$ point, only slightly so
(more at some other points). Thus, even though 
the SO effects are considerable, they are not strong enough to reduce 
the problem to a two   $j_{eff}=1/2$ model.

\begin{figure}[tbh]
\begin{center}
\includegraphics[width=0.9\columnwidth]{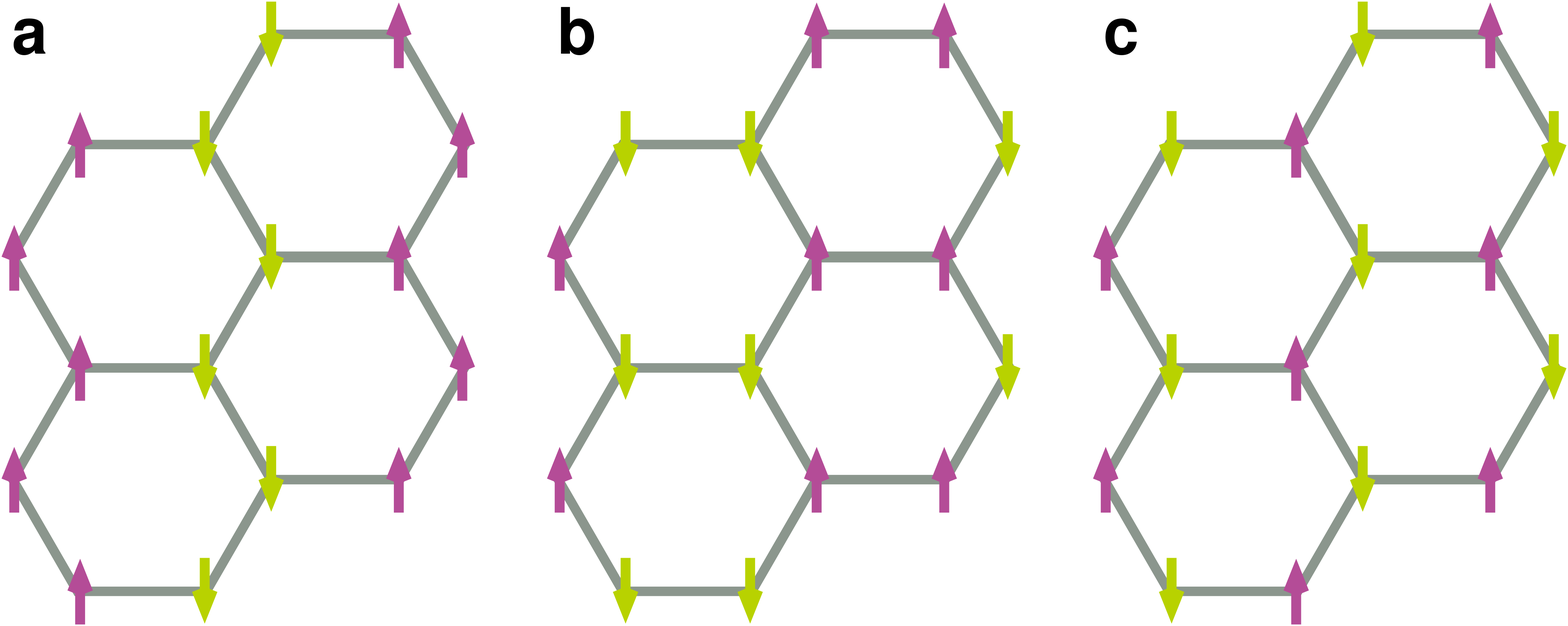}
\end{center}
\caption{ Three antiferromagnetic patterns considered in this paper: (a)
zigzag, (b) stripy, and (c) N\'eel.}
\label{patterns}
\end{figure}

The magnetic patterns considered in our non-relativistic and fully
relativistic calculations are shown in Fig.~\ref{patterns}.

The ferromagnetic state shows in the absence of SO an energy gain of
nearly 80 meV per Ir with respect to the non-magnetic solution and
about half this value against competing antiferromagnetic states
(zigzag and stripy phases); the simple N\'{e}el state is much higher
in energy.  Inclusion of SO changes the energetics considerably, as
described in the main text, with the zigzag antiferromagnetic ordering
becoming competitive with the ferromagnetic one, and lower in energy
than the stripy phase. We deliberately do not discuss the calculated
energies in detail, because the energy differences involved are on
the order of one meV per atom, which is beyond the accuracy of the
density functional theory itself, and on the border of the technical
accuracy of existing band structure codes.

\begin{figure}[tbh]
\begin{center}
\includegraphics[width=0.9\columnwidth]{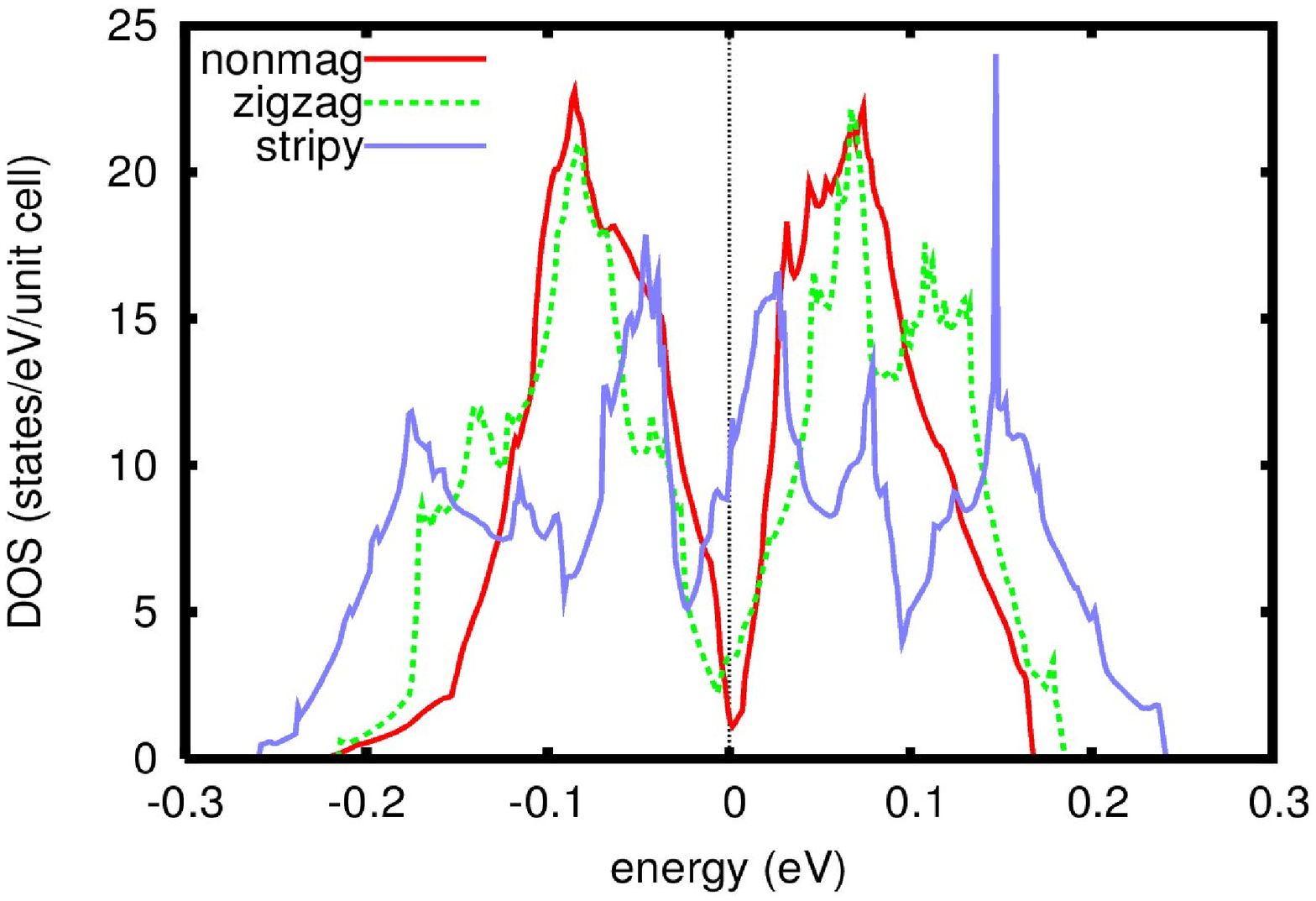}
\end{center}
\caption{Density of states, spin-orbit included, for two competing magnetic patterns compared with
that for the nonmagnetic state. }
\label{DOSs}
\end{figure}

In Fig.~\ref{DOSs} we show the density of states for some magnetic
orderings considered in our fully relativistic calculations. Note that
the zigzag ordering preserves the nonmagnetic pseudogap at the Fermi
level, while the stripy ordering destroys it.

Finally some considerations about the Hubbard $U$ are at place.  In
fact, there are two ways of defining $U$ in this case. As usually, the
actual value of $U$ depends on which orbitals it is being applied
to. For instance, it is well known that in Fe pnictides the
appropriate value of $U$ acting on the Wannier functions combining Fe
$d$ and As $p$ states is more than twice smaller that that acting on
actual atomic $d$ orbitals since the screening effects change
depending on the basis of active states considered. In molecular
solids, such as fullerides, the atomic value of $U$ often appears
completely irrelevant, and the physically meaningful value of $U$ is
the (much smaller) energy of Coulomb repulsion of two electrons placed
on two molecular orbitals. In the case of {\na} one has a
choice of using an atomic $U\sim$ 1.5-2 eV, realizing that the results
will be strongly affected by the fact that electrons are localized not
on individual ions, but on individual QMOs, or of constructing $U$ in
the QMO basis.  The former way is readily available in such formalisms
as LDA+U but it may be a poor choice for the description of a system
based on quasi-molecular orbitals.


\begin{thebibliography}{99}
\bibitem {note1}Note that in this case the spin magnetic moment is 1/3
$\mu_{\rm B}$ and the orbital moment is 2/3 $\mu_{\rm B}.$
\bibitem{gain} Incidentally, the ferromagnetic state in relativistic
  calculations acquires substantial anisotropy.  Depending on the
  polarization direction, the energy gain over the nonmagnetic state
  varies by a factor of two. This is consistent with the
  experimentally observed substantial anisotropy of the uniform
  susceptibility in the paramagnetic state~\cite{gegenwart}.
\end{thebibliography}
\end{document}